\documentclass[aps,10pt,prl,
twocolumn,amssymb,amsmath,amsfonts,
showpacs,floatfix,citeautoscript,preprintnumbers,
footinbib,superscriptaddress]{revtex4-1}
\usepackage{graphicx}
\usepackage{times}
\usepackage{graphicx}
\usepackage[latin1]{inputenc}

\begin{document}

\newcommand{\ket}[1]{\vert{#1}\rangle}
\newcommand{\bra}[1]{\langle{#1}\vert}
\newcommand{\braket}[1]{\langle{#1}\rangle}
\newcommand{\ad}{a^\dagger}
\newcommand{\e}{\ensuremath{\mathrm{e}}}
\newcommand{\norm}[1]{\ensuremath{|{#1}|}}
\newcommand{\aver}[1]{\ensuremath{\big<{#1}\big>}}
\renewcommand{\Im}{\operatorname{Im}}
\newcommand{\etal}{\textit{et al.}}
\newcommand{\moy}[1]{\langle{#1}\rangle}
\newcommand{\elem}[3]{\langle{#1}\vert{#2}\vert{#3}\rangle}
\newcommand{\md}[1]{\vert{#1}\vert}
\newcommand{\dU}{\delta U}
\newcommand{\comment}[1]{{\bf #1}}

\title{Slow quench dynamics of a trapped one-dimensional Bose gas confined to an optical lattice}
\author{Jean-S\'ebastien Bernier}
\affiliation{Centre de Physique Th\'eorique, CNRS, \'Ecole Polytechnique, 91128 Palaiseau Cedex, France.}
\author{Guillaume Roux}
\affiliation{LPTMS, Universit\'e Paris-Sud, CNRS, UMR 8626, 91405 Orsay, France.}
\author{Corinna Kollath$^{1,}$}
\affiliation{Département de Physique Théorique, University of Geneva, CH-1211 Geneva, Switzerland}

\begin{abstract}
  We analyze the effect of a linear time-variation of the interaction
  strength on a trapped one-dimensional Bose gas confined to an
  optical lattice. The evolution of different observables such as the
  experimentally accessible onsite particle distribution are studied
  as a function of the ramp time using time-dependent exact
  diagonalization and density-matrix renormalization group
  techniques. We find that the dynamics of a trapped system typically
  display two regimes: for long ramp times, the dynamics are governed
  by density redistribution, while at short ramp times, local dynamics
  dominate as the evolution is identical to that of an homogeneous
  system. In the homogeneous limit, we also discuss the non-trivial
  scaling of the energy absorbed with the ramp time.
\end{abstract}


\maketitle

Manipulating many-body quantum systems by time-varying their control
parameters is a practical challenge of technological importance in
many areas of physics including condensed matter, quantum information,
and cold atomic and molecular gases.  However, our understanding of
the quantum dynamics of many-particle systems and the identification
of their universal dynamical features is still in its
infancy~\cite{Dziarmaga2009, Polkovnikov2010}. In recent years, it was
suggested that the Kibble-Zurek mechanism~\cite{KibbleZurek},
originally developed to describe the evolution of the early universe,
could explain the dynamics of systems across quantum phase
transitions. Despite a few successes, the validity of this theory to
describe the evolution of all quantum systems is still not accepted.
Unbiased theoretical methods, going beyond scaling arguments, are
required to understand the dynamics of both homogeneous and
inhomogeneous non-integrable quantum systems. In an attempt to shed
some light on the evolution of bosonic systems subjected to a change
of their control parameters, experiments on $^4$He~\cite{Hendry1998}
and more recently on cold atoms were reported. Quenches,
conducted on a trapped bosonic quantum gas loaded into an optical
lattice, were performed by changing the depth of the lattice over a
given time interval~\cite{Greiner2002, Hung2010, Sherson2010,
  Bakr2010}.  In these experiments, the out-of-equilibrium processes
were investigated by considering the behavior of local observables
such as the density, compressibility and onsite particle
distribution. For slow to moderate quenches, two different evolution
regimes which depend on the ramp time and on the initial interaction
strength were observed. These two regimes are believed to be related
to the local and global dynamics of the system~\cite{Hung2010,
  Bakr2010, Rapp2010}.  How well these experiments can be used to
clarify the universal dynamics of homogeneous systems has not been
addressed yet.

In this work, we provide answers to this question by analyzing the
response of bosons stored in a one-dimensional (1D) lattice to a slow
increase of their interaction strength using the unbiased methods of
exact diagonalization (ED) and density-matrix renormalization group
(DMRG). In the homogeneous case, we find, in addition to the sudden
quench and quasi-adiabatic behaviors observed for fast and slow ramp
times respectively, that at intermediate ramp times the absorbed
energy scales non-trivially with the ramp duration.  In the presence
of a trap, we identify two distinct regimes as a function of the ramp
time. For long ramp times, the evolution is governed by density
redistribution, whereas for shorter ramp times, the evolution is
dominated by intrinsic local dynamics and mass transport is
absent. This last response is the same as that of an homogeneous
system.

We carry out our study using the 1D Bose-Hubbard model:
\begin{equation*}
\mathcal{H} = -J \sum_j [b^{\dag}_{j+1} b_j + \text{h.c.}]
        + \frac{U(t)}{2} \sum_j n_j(n_j-1) -\sum_j \mu_j n_j\,,
\end{equation*}
with $b^{\dag}_j$ the operator creating a boson at site $j$, $n_j
=b^{\dag}_j b_j$ the density operator and $J$ and $U$ the hopping and
interaction amplitudes. The chemical potentials $\mu_j$ account for an
external confinement.
At commensurate fillings, a quantum phase transition from a gapless
superfluid (SF) to a gapped Mott insulating (MI) state occurs (at
$U\approx 3.3J$ for $n=1$~\cite{Kuhner2000}).
The slow quench is performed by increasing the interaction strength
linearly, i.e.  $U(t) = U_i + \frac{t}{\tau}\dU$ with $\tau$ the
\emph{ramp time}, $\dU = U_f-U_i$ the quench amplitude, and $U_{i(f)}$
the initial (final) interaction. This can be achieved experimentally
using a suitable Feshbach
resonance~\cite{Inouye1998}. 
Aspects of linear quenches have been discussed previously using
various approximate methods~\cite{Dziarmaga2009, Polkovnikov2010,
  Sengupta2010}. Here, time evolution is computed numerically on
chains of size $L$, using both ED with periodic boundary conditions
and an onsite boson cutoff $M\geq7$, and DMRG~\cite{Daley2004} with
open boundary conditions and $M=6$. The convergence of the DMRG
results both with the number of states (a few hundreds) of the reduced
space and the time-step of the Trotter-Suzuki time-evolution were
checked. Denoting by $E_{0,i/f}$ the initial and final ground-state
energies and $E_f$ the energy obtained at time $\tau$, we introduce
the heat as the energy absorbed by the system: $Q=E_f-E_{0,f}$. Note
that we only consider the time-evolution during the ramp and not the
relaxation once the ramp is completed. 
Further, the derivative of the
chosen ramp has a discontinuity at the initial time, by
which higher modes might be excited.

\begin{figure}[t]
\centering
\includegraphics[width=\columnwidth,clip]{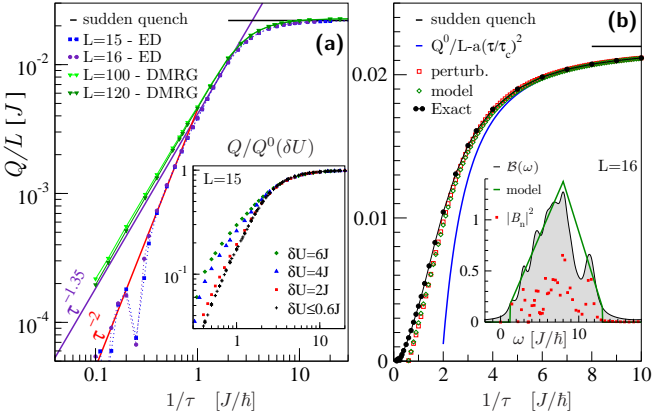}
\caption{(Color online) $U_i=2J$ and $U_f=3J$ at
  $n=1$. \textbf{\textsf{(a)}} Scaling of the heat put into the
  homogeneous system vs inverse ramp time $\tau$ showing a non-trivial
  exponent in the near-adiabatic regime. \textit{Inset} : collapse of
  data for different $\dU$ (ED). \textbf{\textsf{(b)}} Comparison to
  perturbation theory ($\tau_c \simeq 3.47\hbar/J$). \textit{Inset}: 
  Spectral function associated with perturbation theory for $U=2J$ (ED).}
 \label{fig:homsys-heat}
\end{figure}

\emph{Homogeneous system} -- We first aim here at understanding the
intrinsic evolution of local observables by studying the homogeneous
limit as many features identified in such systems are also relevant to
the trapped case. In Fig.~\ref{fig:homsys-heat}(a), we show the heat
produced by the quench for a given $\dU$ as a function of the inverse
ramp time $\tau^{-1}$. In finite-size systems, three regimes are
typically observed: at large-$\tau$ (adiabatic limit), $Q \sim
\tau^{-2}$ with oscillations~\cite{Dziarmaga2009}, here associated
with a finite-size gap~\cite{FootNote1};
at intermediate $\tau$, we find a non-trivial power-law behavior;
finally, for short $\tau$, $Q$ approaches the sudden quench
limit~\cite{Kollath2007} quadratically.

These results can be compared with time-dependent perturbation theory
in the \emph{initial} Hamiltonian basis
$\{\ket{n}\}$ 
where the ``ramp velocity'' $v=\dU/\tau$ is the small parameter. To
second order in $v$, the energy variation reads:
\begin{equation*}
\label{eq:pertE}
\Delta E(t) =  vt B_0 - v^2 \sum_{n \neq 0}
 \frac{\md{B_n}^2}{\omega_n^3}
\left[(\omega_n t)^2 - 4\sin^2\left(\frac{\omega_nt}{2}\right)\right]
\end{equation*}
with $\hat{B}=\sum_j n_j(n_j-1)/2$ the perturbation operator and
$B_n=\elem{n}{\hat{B}}{0}$. Numerically, we access the $\md{B_n}^2$
and energies $\omega_n=E_n-E_0$ in the low-energy regime using the
Lanczos algorithm. Considering only the final time $\tau$ and
introducing the energy spectral function, $\mathcal{B}(\omega) =
\sum_n \md{B_n}^2 \delta(\omega-\omega_n)$, one gets for the heat
\begin{eqnarray}
\label{eq:pertQ}
  Q(\tau,\dU) &=& Q^{0}(\dU)\\
  \nonumber
  &-&\frac{\dU^2}{\tau ^2}\!\int_{0^+}^{\infty}\!\!d\omega
\frac{\mathcal{B}(\omega)}{\omega^3} \left[(\omega\tau)^2\!-4\sin^2\!\left(\frac{\omega \tau}{2}\right)\right],
\end{eqnarray}
where $Q^{0}(\dU) = \dU B_0+E_{0,i}-E_{0,f}$ is the \emph{exact}
sudden quench expectation. Counter-intuitively, even though this
result is perturbative in $v$, it yields the correct result in the
large velocity $v$ limit as the ramp time $\tau$ becomes short enough
in this regime. Thus, this perturbation theory provides short-$\tau$
corrections away from the sudden quench limit, which can be calculated
using ground-state observables. Indeed, assuming that
$\mathcal{B}(\omega)$ has a support with an upper bound or decays
exponentially, ones finds that $Q(\tau,\dU) = Q^{0}(\dU) -
L\frac{\dU^2}{J}(\tau/\tau_c)^2$ with the characteristic ramp time
$\tau_c$ given by
\begin{equation*}
\tau_c^{-2} = \frac{J}{12L} \int_{0}^{\infty}\!\!d\omega\,\omega\mathcal{B}(\omega)
            = \frac{J^2}{12L}\elem{0}{B[B,K]}{0}
\end{equation*}
where $K$ is the kinetic term $\sum_j [b^{\dag}_{j+1} b_j +
\text{h.c.}]$. The magnitude of $\tau_c$ is consequently directly
connected to the equilibrium three-point correlator $B[B,K]$. The
scaling of $\tau_c$ with $L$ depends on the typical behavior of the
correlator with distance. 
In our model, we find with
negligible finite-size effects that $\tau_c$ is an increasing function
of $U_i$, indicating that the correlator drops off rapidly for any
$U$. 
Using Eq.~(\ref{eq:pertQ}), one obtains a quantitative agreement up to
intermediate velocities and close to the power-law regime (see
Fig.~\ref{fig:homsys-heat}(b)). Furthermore, the details of
$\mathcal{B}(\omega)$ do not alter much this short-$\tau$ regime as a
truncated triangular approximation of $\mathcal{B}(\omega)$ (``model''
in Fig.~\ref{fig:homsys-heat}(b)) reproduces well this
regime.

We observe that the accuracy of Eq.~(\ref{eq:pertQ}) improves 
with decreasing $\dU$. In the small $\dU$ limit, the quench
essentially becomes a probe of the initial ground-state dynamics. To
second order in $\dU$, the maximum heat behaves as $Q^{0}(\dU) \simeq
\dU^2 \sum_{n>0}\md{B_n}^2/\omega_n$. Combining this result with
Eq.~(\ref{eq:pertQ}), we find that $Q(\tau,\dU)/Q^{0}(\dU) \simeq
f(\tau)$ is a function of $\tau$ only
. In the Inset of
Fig.~\ref{fig:homsys-heat}(a), we see that curves do collapse well
onto each other for $\dU \leq 0.6$; while they do not for larger $\dU$
as in this case heat production depends on both $\tau$ and $\dU$.
Remarkably, the small $\dU$ limit also provides
information on the large-$\tau$ (adiabatic) regimes as, in that case,
the first part of the integral in Eq.~(\ref{eq:pertQ}) cancels with
$Q^{0}(\dU)$. There, we recover the results of
Ref.~\onlinecite{Eckstein2010}. 
In
particular, we see that for a gapped spectrum Eq.~(\ref{eq:pertQ})
reproduces the typical $\tau^{-2}$ decay combined with some
oscillating terms. 
Lastly, we discuss the exponent $\eta\approx 1.35$ observed for
$U_i=2J$ and $\dU=J$. For large system sizes, this behavior is found
above the sudden quench limit but below a ramp time set by the inverse
finite-size gap~\cite{Canovi2009}. We cannot relate this exponent to
various predictions on approximated versions of the
model~\cite{Dziarmaga2009}, nor to the above perturbative
approach. Similar non-universal behaviors were recently reported in
Ref.~\onlinecite{Canovi2009}. Moreover, the Inset of
Fig.~\ref{fig:homsys-heat}(a) shows that this exponent depends on
$\dU$. This may have different reasons. First, contributions of
intermediate interaction values at which the system possibly possesses
different low-energy physics might play a role. Second, a
discontinuity in the derivative of the considered ramp form might
cause higher energy excitations.  Last, energy redistribution
between excited states, which we found to be negligible only for small
quenches, could also play a role. This exemplifies that, typically,
only the small $\dU$ limit can display universality.

\begin{figure}[t]
\centering
\includegraphics[width=0.95\columnwidth]{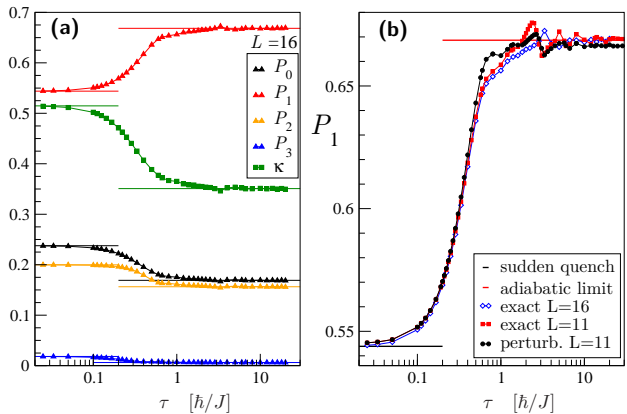}
\caption{(Color online) Observables for a quench from $U_i=2J$ to
  $U_f=4J$ (ED) \textbf{\textsf{(a)}} Occupation probability $P_n$ and
  compressibility $\kappa$ vs ramp time $\tau$ in the homogeneous
  system. \textbf{\textsf{(b)}} Perturbation theory of
  Eq.~(\ref{eq:pert-A}) gives a quantitative prediction for
  $L=11$. The $L=16$ data gives an idea of finite-size effects.}
\label{fig:homsys-P1}
\end{figure}

Following experiments, we turn to the discussion of local
observables. $\kappa_j = \moy{n_j^2}-\moy{n_j}^2$ denotes the
compressibility at site $j$ while $P_n$ is the probability of having
$n$ bosons onsite.
The evolution of Fig.~\ref{fig:homsys-P1}(a) starting from the SF phase
typically displays the same three regimes as for the heat, with
finite-size oscillations close to the adiabatic limit. Entering the MI
at $n=1$, $P_1$ increases with $U$ while other $P_{n}$ concomitantly
decrease. The perturbative prediction for a real symmetric observable
$A$ reads, at time $\tau$ and to first order in $v$,
\begin{equation}
\label{eq:pert-A}
A(\tau,\dU) = A_0
- 2 \frac{\dU}{\tau} \sum_{n\neq 0} 
\frac{\omega_n\tau-\sin(\omega_n\tau)}{\omega_n^2} A_nB_n\;.
\end{equation}
Here again, it is remarkable that the simple perturbation
theory provides a quantitatively good (up to a percent in this case)
prediction of the exact time-evolution
(see Fig.~\ref{fig:homsys-P1}(b)).

\emph{Trapped system} -- In current cold atom experiments, an external
parabolic trap is usually present, i.e.~$\mu_j = -V(j-\frac{L+1}{2})^2$. 
As the density increases from the edges to the center, for large $U$,
MI domains coexist with SF regions~\cite{Batrouni2002}. Thanks to
recent advances, measurements with single-site resolution are now
possible in 2D cold atoms lattice setups~\cite{Bakr2010,Sherson2010},
enabling one to focus on a particular domain. However, in an
out-of-equilibrium situation, density gradients and parameter changes
induce flows of particles. Consequently, the density redistribution
will have an impact on locally measured quantities. These transport
phenomena occur on timescales that depend on the velocity of
excitations (typically controlled by $J$) and the domain sizes.

\begin{figure}[t]
\centering
\includegraphics[width=0.95\columnwidth,clip]{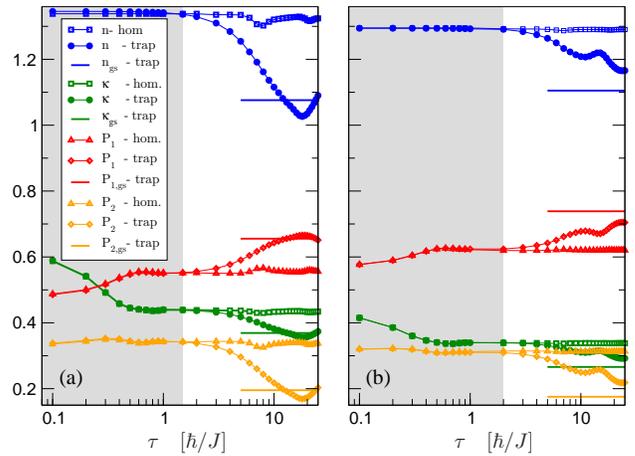}
\caption{(Color online). Comparing local observables at the center of
  the system between homogeneous and trapped ($V=0.006J$) configurations as a
  function of the ramp time $\tau$ (DMRG calculations). The grey area
  shows the timescales over which the responses are
  identical. \textbf{\textsf{(a)}} slow change from $U_i=2J$ to
  $U_f=4J$. \textbf{\textsf{(b)}} $U_i=4J$ to $U_f=6J$.}
\label{fig:trapsys_pofn}
\end{figure}

Generally, the timescales for density redistribution are longer than
the intrinsic dynamics of a local observable. We show in this section
that, in the fast quench regime, an intrinsic evolution of the
observables occurs while the density redistribution remains frozen. On
the contrary, this redistribution dominates in the slow ramp
regime. To demonstrate this point, we systematically compare the
evolution of observables in the trapped cloud with its homogeneous
counterpart (choosing the same initial density in the center). We
consider the evolution of a trapped cloud whose majority of atoms are
initially in a SF state as the system features no or only weak MI
regions signaled by a trough in the
compressibility. Fig.~\ref{fig:trapsys_pofn} displays comparisons
between observables taken at the central site as a function of the
ramp time, and for two different interaction quenches: (a)
$U=(2\rightarrow4)J$, (b) $U=(4\rightarrow6)J$.

\begin{figure*}[t]
\centering
\includegraphics[height=4.45cm,clip]{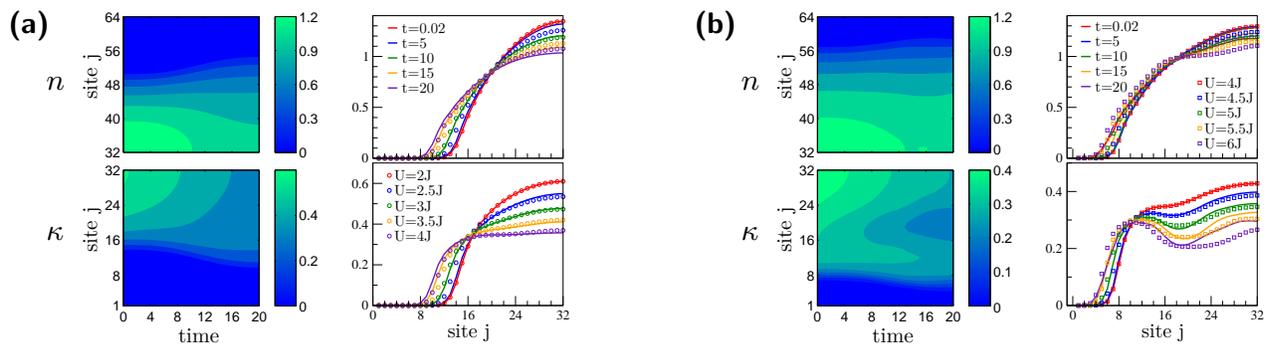}
\caption{Slow quenches with $\tau=20\hbar/J$, $L=64$ and $V=0.006J$.
  \textbf{\textsf{(a)}} $U=(2\rightarrow 4)J$, \textbf{\textsf{(b)}}
  $U=(4\rightarrow 6)J$. Contour plots: density $n$ (up) and
  compressibility $\kappa$ (down) as a function of position and
  time. Graphs: $n$ (up) and $\kappa$ (down) vs position for various
  times $t$ (full lines). Circles and squares stand for ground-state
  predictions with corresponding $U(t)$.}
\label{fig:contourmap}
\end{figure*}

For case (a) of Fig.~\ref{fig:trapsys_pofn}, we see that for short
ramp times the central density does not rearrange. In fact, the
density remains constant for all ramp times shorter than $\tau
\approx\hbar/J$. Meanwhile, the compressibility $\kappa$ and
probabilities $P_{1/2}$ show well pronounced variations. Remarkably,
for this range of $\tau$, we find all observables to be in excellent
agreement with the homogeneous system predictions. This result
supports our previous statement that intrinsic local dynamics dominate
at short ramp times. For longer ramp times $\tau>2\hbar/J$, we find a
clear change in the central density, associated with the onset of
particle currents~\cite{FootNote3}.
Naturally, this reduction of the density modifies the particle
distribution and compressibility, driving them away from the constant
density behavior and close to the adiabatic expectation for the
trapped cloud (with slight oscillations). On
Fig.~\ref{fig:contourmap}(a), we show the actual time evolution of the
density and compressibility corresponding to a ramp time
$\tau=20\hbar/J$. The density configuration remains almost frozen up
to time $t=5\hbar/J$ even though the corresponding ground state
considerably differs, and begins to evolve afterward. At later times,
the density profile broadens reaching a wider size than the one
expected for the ground state at the corresponding $U/J$. At the end
of the evolution process, the system is therefore in an excited state.
In comparison, the compressibility distribution evolves much faster
than the density. At each time step, the compressibility is relatively
close to its corresponding ground state value. The difference between
the time-evolved and ground-state values are most likely due to the
unrelaxed density profile. Consequently, this direct time-analysis of
a slow quench reinforces our previous conclusion that the intrinsic
evolution of the observables and the density redistribution are
characterized by two different timescales.

For case (b) of Fig.~\ref{fig:trapsys_pofn}, the first striking
feature is the even slower evolution of the central density with
increasing ramp times when compared to the situation discussed
before. Up to $\tau \approx 2 \hbar/J$, the density remains very close
to its initial value and the other local observables are identical to
their corresponding homogeneous value. For longer ramp times, the
central density then shows an oscillatory decrease and the other
observables deviate from their constant density counterparts. However,
even for the longest ramp times considered, the density has still not
reached the adiabatic regime. The slowdown of the density
rearrangement is attributed to the emergence of Mott ``barriers'' in
regions where $n_j\approx 1$. This effect is underlined in the
snapshots of Fig.~\ref{fig:contourmap}(b). These ``barriers'' arise
due to the rapid reduction of particle fluctuations at large $U$ and
the local reduction of transport. On Fig.~\ref{fig:contourmap}(b), we
observe a much slower redistribution of the density than for the
$U_i=(2\rightarrow4)J$ evolution in (a). In this case, the density
profile is actually narrower than the ground state profile. The
compressibility presents the same fast evolution at short time,
building ``barriers'' before the currents can establish and precluding
the redistribution from taking place.

In summary, we performed a detailed study of slow quenches in the
trapped 1D Bose-Hubbard model. We identified two dynamical
regimes. For short ramp times during which currents cannot develop,
the intrinsic local dynamics dominates and the response is equivalent
to that of an homogeneous system.  Many features of this regime are
well described by perturbation theory. For longer ramp times, currents
do set in and the dynamics is governed by non-trivial transport
phenomena. For large final interaction strengths, the global
timescales are significantly enhanced by the presence of ``Mott
barriers''. Our results agree qualitatively well with recent
experiments focusing on transport dynamics of cold atoms across the
SF--MI transition~\cite{Hung2010, Bakr2010}. Indeed, based on our
results, we can argue that the two experiments were probably conducted
in different regimes. In Ref.~\onlinecite{Hung2010}, the slowdown of
dynamics, due to the presence of Mott regions, indicates that their
experiment was carried out in the regime with global dynamics.  In
contrast, the short timescales observed in Ref.~\onlinecite{Bakr2010}
are most likely related to local dynamics.

\emph{Note added}: During the final stages of the preparation of this
manuscript, a mean-field study of the 2D version of the model
appeared on arXiv~\cite{Natu2010}.

\emph{Acknowledgments} -- We acknowledge discussions with 
P.~Barmettler, G.~Biroli, C.~De Grandi, A.~Silva and P.~Werner.
Support was provided by ANR (FAMOUS), DARPA-OLE, Triangle de la 
physique and FQRNT.


\begin{thebibliography}{10}
\bibitem{FootNote1}
A finite-size gap $\propto 1/L$ is present even in the SF phase.
\bibitem{FootNote3}
The density oscillations observed in the homogeneous system at large $\tau$ 
are due to boundary effects.
\end{thebibliography}
\end{document}